\documentclass{article}
\def\bea{\begin{eqnarray}}
\def\eea{\end{eqnarray}}

\def\rl{\rho_1}
\def\rr{\rho_2}

\def\ot{\otimes}
\def\one{\hbox{{\small $1$} \hskip -6.7pt {\normalsize $1$}}}
\def\dis{\displaystyle}

\def\a{A_}
\def\b{B_}

\usepackage{amsmath}
\usepackage{cite}

\begin{document}
\begin{titlepage}
\begin{center}
{\large \textbf{Multi-species reaction-diffusion models admitting
shock solutions}} \vskip 2\baselineskip \centerline {\sffamily S.
Masoomeh Hashemi\footnote{e-mail: hashemy.m@gmail.com} \&
 Amir Aghamohammadi\footnote
 {e-mail: mohamadi@alzahra.ac.ir}}
 \vskip 2\baselineskip
{\it Department of Physics, Alzahra University, Tehran 19384,
IRAN}
\end{center}


\begin{abstract}
\noindent A method for classifying $n$-species reaction-diffusion
models, admitting shock solutions is presented. The most general
one-dimensional two-species reaction-diffusion model with nearest
neighbor interactions admitting uniform product measures as the
stationary states is studied. Satisfying more constraints, these
models may experience single-shock solutions. These models are
generalized to multi-species models. The two-species models are
studied in detail. Dynamical phase transitions of such models are
also investigated.

\end{abstract}
\end{titlepage}
\newpage
\section{Introduction}
The stochastic modeling of systems is a useful method for studying the
problems in non-equilibrium statistical physics. $\o$
Reaction-diffusion systems are stochastic models which can be used
to study the evolution of interacting particle systems. Extensive
researches have been done on one-dimensional reaction-diffusion
systems, some of which belong to the emergence and evolution of
shocks whose positions perform random walks, i.e. density
discontinuities which are randomly on the move. The simplicity of
the asymmetric simple exclusion process, including just diffusive
processes, provided researchers with a suitable ground to take
first steps in exploring the dynamics of shocks
\cite{DLS,F,FFV,BES,B}. In analogy to ASEP some interesting models
have been introduced, for example a driven diffusive two-channel
system \cite{POS} or bricklayers' model, which is a model without
exclusion but yet uncorrelated \cite{BA,TV}. To take into account
the systems including interacting processes, for instance, shock
formation in driven diffusive systems, containing homogeneous
creation and annihilation of particles has been investigated
\cite{PRWKS,EJS} and recently more complicated systems have been
described \cite{PS,RS,TS1,TS2,JM}.

It is known that the ordinary Glauber model on a one-dimensional
lattice with boundaries at any temperature, shows a dynamical
phase transition \cite{KA01}. The dynamical phase transition is
controlled by the rate of spin flip at the boundaries, and is a
discontinuous change of the derivative of the relaxation time
towards the stationary configuration. In \cite{KA02}, using a
transfer matrix method, it is shown that a one-dimensional kinetic
Ising model with nonuniform coupling constants may exhibits a
dynamical phase transition. Other phase transitions induced by
boundary conditions have also been studied ( \cite{HGS,RiK,KA03}
for example).

We are interested in the works aim to introduce new solvable
models. In \cite{KJS} a single-species model with nearest neighbor
interactions on a one dimensional lattice with open boundaries has
been considered. It was shown that there are three families of
models with traveling wave solutions; the asymmetric simple
exclusion process (ASEP), the branching-coalescing random walk
(BCRW) and the asymmetric Kawasaki-Glauber process (AKGP). A
classification of single-species models with three-site
interactions and  special choice of symmetries has been studied in
\cite{PS}. Recently some efforts have been made to obtain the
models on a lattice with two types of particles and nearest
neighbor interactions. In \cite{RS} a model with diffusion and
exchange processes has been studied. In \cite{TS1} a model with a
degenerate conservation law and PT invariance (invariant under the
application of time reversal and space reflection) has been
discussed. Another class of two-species models with a
non-degenerate conservation law has been presented in \cite{TS2}.
Besides the straightforward calculation of the master equation
there is an alternative approach in dealing with
reaction-diffusion problems which is the so-called matrix product
formalism. It is an algebraic method that takes advantage of
non-commutative operators instead of probabilities. In \cite{JM}
it is assumed that the density of A particles in a given site is
proportional to the density of B particles at the same site. Using
matrix product formalism, they have found three three-states
models which are basically two-state systems. One of these models
is the generalization of \cite{TS1}.

In this article there is an attempt to present a method for
classifying $n$-species particle systems, admitting single shock
solutions. In section 2, after a brief review of formalism, a
two-species reaction-diffusion model on a one-dimensional lattice
with boundaries is introduced. In section 3, the most general
two-species models with nearest neighbor interactions, admitting
uniform product measures as the stationary states are studied. For
such models reaction rates should satisfy some constraints. These
conditions are obtained. In section 4, a single-shock measure is
introduced and a classification method is presented. Then, it is
generalized to multi-species models. In section 5, dynamical phase
transitions of some two-species models are investigated. Finally, a
discussion on the results of this article and those of previous
ones is presented.

\section {Formulation}
Two-species reaction-diffusion models with nearest-neighbor
interactions on a one-dimensional lattice with $L$ sites are
studied. There are two types of particles denoted by A and B. The
processes are exclusive which means each site is either empty
(will be shown by $\emptyset$) or occupied by at most one
particle, A or B. The empty state is denoted by $| \emptyset
\rangle$, and the occupied state by particle A (B) is represented
by $|{\rm A}\rangle$ ($|{\rm B}\rangle$). The 3-dimensional vector
space for each site is spanned by
\begin{eqnarray}
| \emptyset \rangle=\left(\begin{array}[c]{c}1\\0\\0\end{array}\right),\qquad
|{\rm A}\rangle=\left(\begin{array}[c]{c}0\\1\\0\end{array}\right),\qquad
|{\rm B}\rangle=\left(\begin{array}[c]{c}0\\0\\1\end{array}\right).
\end{eqnarray}
The $3^L$-dimensional
vector space of the lattice is given by the tensor product of the
single-site vector spaces $V^3$
\begin{eqnarray}
V=\underbrace{V^3\ot\cdots\ot V^3}_{L}.
\end{eqnarray}
The state vector of the system is
\begin{eqnarray}
|P\rangle_t=\sum_{\eta=1}^{3^L} P(\eta,t)|\eta\rangle,
\end{eqnarray}
where $|\eta\rangle$ is the basis vector of the lattice and
$P(\eta,t)$ is the probability of finding the system in state
$\eta$ in time $t$. Continuous-time Markovian evolution is the
master equation
\begin{eqnarray}
\frac{d|P\rangle_t}{dt}=H|P\rangle_t.
\end{eqnarray}
The non-diagonal elements of the generator $H$ are the transition
rates $\omega_{ji}$, the transition rate from state $i$ to $j$, and
its diagonal elements are the negative sum of the non-diagonal
elements of their own columns. So the sum of each column is zero,
and consequently
\begin{eqnarray}
\langle S|H=0,
\end{eqnarray}
where $\langle S|$ is a row vector whose all elements are equal to one. Let us denote  $\omega_{+i}:=\sum_{j\neq i}\omega_{ji}$.\\

For a two-species model the local hamiltonian $h_{k,k+1}$ is a
$9\times 9$ matrix, which acts on sites $k$ and $k+1$, and
contains $72$ two-site transition rates. If the single-site states
are respectively assumed empty ($\emptyset$), occupied by A and B,
the two-site states respectively become:
\begin{eqnarray}
1)\ \emptyset\emptyset\qquad \qquad \ & 4)\  {\rm A}\emptyset
\qquad \qquad& 7)\  {\rm B} \emptyset\cr 2)\ \emptyset {\rm
A}\qquad \qquad & 5)\  {\rm A}{\rm A}\qquad \qquad& 8)\  {\rm
B}{\rm A}\cr 3)\ \emptyset{\rm B} \qquad \qquad & 6)\  {\rm A}{\rm
B}\qquad \qquad& 9)\  {\rm B}{\rm B}
\end{eqnarray}
Let us assume open boundary conditions, where the particles can
enter and leave the lattice from the first and last sites with the
following rates
\begin{eqnarray}
\alpha_{12}:\ {\rm A}\rightarrow \emptyset,&\alpha_{21}:\ \emptyset\rightarrow {\rm A}\cr
\alpha_{13}:\ {\rm B}\rightarrow \emptyset,&\alpha_{31}:\ \emptyset\rightarrow {\rm B}
\end{eqnarray}
and change to each other with the rates
\begin{eqnarray}
\alpha_{23}:\ {\rm B}\rightarrow {\rm A}, &\alpha_{32}:\ {\rm A}\rightarrow {\rm B}
\end{eqnarray}
The hamiltonians $B_1$ and $B_L$ are for the left and right
boundaries and  are in the form
\begin{eqnarray}
B_{1/L}=\begin{pmatrix}-(\alpha_{21}^{\ell/r}+\alpha_{31}^{\ell/r})&\alpha_{12}^{\ell/r}&\alpha_{13}^{\ell/r}\\
 \alpha_{21}^{\ell/r}&-(\alpha_{12}^{\ell/r}+\alpha_{32}^{\ell/r})&\alpha_{23}^{\ell/r}\\
\alpha_{31}^{\ell/r}&\alpha_{32}^{\ell/r}&-(\alpha_{13}^{\ell/r}+\alpha_{23}^{\ell/r})\end{pmatrix},
\end{eqnarray}
where the indices $\ell$ and $r$ stand for left and right. So the
hamiltonian for a one-dimensional $L$-site lattice with
nearest-neighbor interactions and single-site interactions at the
boundaries is in the following  form
\begin{eqnarray}\label{hamiltinian}
H=B_1\ot\one^{\ot(L-1)}+\sum_{k=1}^{L-1}\one^{\ot(k-1)}\ot
h_{k,k+1}\ot\one^{\ot(L-k-1)}+\one^{\ot(L-1)}\ot B_L
\end{eqnarray}
 $\one$
denotes a $3\times 3$ unit matrix. We assume the interactions to
be homogenous, so $h_{k,k+1}$ is the same for all sites. We also
assume the interactions to be time independent.

\section { Uniform product measures as stationary states}
Let us consider the state vector of the system  to be
uncorrelated. Then the state of the system is the tensor product
of the single-site state vectors
\begin{equation}
|P\rangle = |P_1\rangle\ot |P_2\rangle\ot\cdots\ot |P_L\rangle\
\end{equation}
where $|P_k\rangle$ is the state vector of site $k$. If the
occupation probability of particle ${\rm A}$ in site $k$ is $a_k$
and the occupation probability of particle ${\rm B}$ in site $k$
is $b_k$, the probability of being empty becomes $(1-(a_k+b_k))$.
Consequently
\begin{equation}
|P_k\rangle=\left(\begin{array}[c]{c}1-(a_k+b_k)\\a_k\\b_k\end{array}\right)
\end{equation}
where not only $0\leq a_k\ ,\ b_k\leq 1$ but also $0\leq
a_k+b_k\leq 1$. We also assume that the densities are uniform, and
the state vector of each site is shown by $|u\rangle$. Then the
state vector of the lattice becomes
\begin{equation}\label{station}
|P\rangle =|u\rangle^{\ot L},
\end{equation}
which is a uniform product measure. The state vector $|P\rangle$
is the stationary state of the system if
\begin{equation}\label{station1}
H |P\rangle=\ 0,
\end{equation}
which gives
\begin{equation}\label{hu}
H |u\rangle^{\ot L}=\ 0,
\end{equation}
One can expand $h|u\rangle\ot |u\rangle$ in the following form
\begin{equation}\label{huu}
h|u\rangle\ot |u\rangle\ =A|u\rangle\ot
|u\rangle+\sum_{\mu,\nu=1}^2B_{\mu\nu}|x_\mu\rangle\ot
|x_\nu\rangle+\sum_{\mu=1}^2\{C_\mu|u\rangle\ot
|x_\mu\rangle+D_\mu|x_\mu\rangle\ot |u\rangle\},
\end{equation}
where $A$, $B_{\mu\nu}$, $C_\mu$, and $D_\mu$ are constant, and
the two vectors $|x_1\rangle$ and $|x_2\rangle$ together with
$|u\rangle$ form a linearly independent set. Hence, substituting
the above expansion into the equation (\ref{hu}) leads to a
combination of linearly independent terms. Thus, one deduces
\begin{equation}
A=\ B_{\mu\nu}=0,\qquad C_\mu+D_\mu=0.
\end{equation}
Therefore, for an infinite lattice or a periodic one, where there
is no boundary term, equation (\ref{huu}) recasts to
\begin{equation}
h|u\rangle\ot |u\rangle\ =|u\rangle\ot |x\rangle-|x\rangle\ot |u\rangle,
\end{equation}
where the  vector $|x\rangle$ is a linear combination of
$|x_1\rangle$ and $|x_2\rangle$. This gives some constraints on
the reaction rates. For a lattice with boundaries, $|u\rangle^{\ot
L}$ is a stationary uniform product measure provided that
\begin{eqnarray}
&&h|u\rangle\ot |u\rangle\ =|u\rangle\ot |x\rangle-|x\rangle\ot
|u\rangle,\cr &&B_1 |u\rangle=|x\rangle +\alpha |u\rangle ,\cr
&&B_L  |u\rangle=-|x\rangle -\alpha |u\rangle,
\end{eqnarray}
where $\alpha$ is a constant, which depends on the reaction rates. Defining the vector $|{\cal R}\rangle$ by
\begin{equation}
|{\cal R}\rangle:= h|u\rangle\ot |u\rangle,
\end{equation}
then one obtains
\begin{eqnarray}\label{constraint01}
&R_1=R_5=R_9=0,&R_2+R_4=0,\cr&&\cr &R_3+R_7=0,&R_6+R_8=0,
\end{eqnarray}
and
\begin{equation}\label{constraint02}
\frac{R_2}{u_1u_2}+\frac{R_6}{u_2u_3}=\frac{R_3}{u_1u_3}.
\end{equation}
where $R_i$ is the $i$th element of the vector $|{\cal R}\rangle$.
Defining $\rho$ by
\begin{equation}
\rho:=a+b
\end{equation}
(\ref{constraint01}) may be written as
\begin{eqnarray}
&&-\omega_{+1}(1-\rho)^2+(\omega_{12}+\omega_{14})(1-\rho)a+(\omega_{13}+\omega_{17})(1-\rho)b\cr
&&+(\omega_{16}+\omega_{18})ab+\omega_{15}a^2+\omega_{19}b^2=0\cr\cr
&&\omega_{51}(1-\rho)^2+(\omega_{52}+\omega_{54})(1-\rho)a+(\omega_{53}+\omega_{57})(1-\rho)b\cr
&&+(\omega_{56}+\omega_{58})ab-\omega_{+5}a^2+\omega_{59}b^2=0\cr\cr
&&\omega_{91}(1-\rho)^2+(\omega_{92}+\omega_{94})(1-\rho)a+(\omega_{93}+\omega_{97})(1-\rho)b\cr
&&+(\omega_{96}+\omega_{98})ab+\omega_{95}a^2-\omega_{+9}b^2=0\cr\cr
&&(\omega_{21}+\omega_{41})(1-\rho)^2+(-\omega_{+2}+\omega_{24}+\omega_{42}-\omega_{+4})(1-\rho)a\cr
&&+(\omega_{23}+\omega_{27}+\omega_{43}+\omega_{47})(1-\rho)b+(\omega_{25}+\omega_{45})a^2\cr&&
+(\omega_{26}+\omega_{28}+\omega_{46}+\omega_{48})ab
+(\omega_{29}+\omega_{49})b^2 =0\cr\cr
&&(\omega_{31}+\omega_{71})(1-\rho)^2+(\omega_{32}+\omega_{34}+\omega_{72}+\omega_{74})(1-\rho)a\cr
&&+(-\omega_{+3}+\omega_{37}+\omega_{73}-\omega_{+7})(1-\rho)b+(\omega_{35}+\omega_{75})a^2\cr
&&+(\omega_{36}+\omega_{38}+\omega_{76}+\omega_{78})ab
+(\omega_{39}+\omega_{79})b^2=0\cr\cr
&&(\omega_{61}+\omega_{81})(1-\rho)^2+(\omega_{62}+\omega_{64}+\omega_{82}+\omega_{84})(1-\rho)a\cr
&&+(\omega_{63}+\omega_{67}+\omega_{83}+\omega_{87})(1-\rho)b+(\omega_{65}+\omega_{85})a^2\cr
&&+(-\omega_{+6}+\omega_{68}+\omega_{86}-\omega_{+8})ab +(\omega_{69}+\omega_{89})b^2=0
\end{eqnarray}
These constraints guarantee existence of a uniform product measure
as a stationary state. Depending on the model, the stationary
state may be unique. For such models, for any initial state, this
product measure is the final state.

\section {Product shock measures}
Let us assume  the occupation probability of particle A (B) in the
first $m$ sites is $a_1$ ($b_1$) and in the latter $(L-m)$ sites
is $a_2$ ($b_2$). This is  a single product shock measure, and the
state vector is
\begin{equation}
|e_m\rangle=|u\rangle^{\ot m}\ot |v\rangle^{\ot (L-m)}
\end{equation}
where
\begin{eqnarray}
&|u\rangle=\left(\begin{array}[c]{c}1-\rho_1\\ a_1\\
b_1\end{array}\right),&
|v\rangle=\left(\begin{array}[c]{c}1-\rho_2\\ a_2\\ b_2\end{array}\right).
\end{eqnarray}
We briefly call it a shock measure. The shock state vectors
$|e_m\rangle,\ m=0,1,\cdots ,L$\ form $L+1$ vectors and are a
closed set under time evolution, if the evolution equations are in
the form of
\begin{eqnarray}\label{shock2}
 &&H|e_m\rangle =d_1|e_{m-1}\rangle+d_2|e_{m+1}\rangle-(d_1+d_2)|e_m\rangle, \quad 1\leq m \leq L-1,
\end{eqnarray}
and
\begin{eqnarray}
 &&H |e_0\rangle=D^{'}_1|e_1\rangle-D^{'}_1|e_0\rangle,\cr
 &&H\ |e_L\rangle=D^{'}_2|e_{L-1}\rangle-D^{'}_2|e_L\rangle.
\end{eqnarray}
The shock position can perform one-site jumps with the rate $d_1$
($d_2$) to the left (right) and at the left (right) boundary with
the rate $D'_1$ ($D'_2$) to the right (left). As can be seen, the
initial shock state can evolve into a linear combination of the
shocks. It is said the shocks make an invariant sub-space.
$d_1,d_2,D'_1$, and $D'_2$ are all non-negative and called the
hopping rates of the shock position.

A shock is immobile when all the hopping rates $d_1,d_2,D'_1$ and
$D'_2$ become zero. Then the shock does not move in the lattice,
and the two stationary uniform product measures $|e_0\rangle,
|e_L\rangle$ are stationary.

\subsection{ Classification of shocks containing the occupation probabilities different from 0 and 1}
When there is a shock in the lattice, it is divided into two
parts. The occupation probabilities for particles ${\rm A}$ and
${\rm B}$ in each part are $a_k$ and $b_k$, consequently $0<a_k\
,\ b_k< 1$. We also have  $0< \rho_k:=a_k+b_k< 1$. To have a
shock, at least one of the particle densities in the first part of
the lattice, $a_1$ or $b_1$, should be different from that in the
other part. Here we have assumed that $a_1\ne a_2$, $b_1\ne b_2$,
and $\rl\neq\rr$. A shock measure is constructed of two stationary
uniform product measures, so the reaction rates should satisfy two
sets of the equations (\ref{constraint01},\ref{constraint02}). To
classify the shocks, we should first define equivalent models. One
may obtain models which apparently seem to be different, but there
exist transformations which relate them together. We take these
models as equivalent models.
\begin{itemize}
\item{Under the left and  right exchange, some different models
transform to each other so are equivalent.}
\item{ The models which
change to each other through the transformation
A$\leftrightarrow$B, are equivalent. Besides, because there is not
any difference between the nature of an empty site and a site
occupied by a particle, the empty site can be considered as a
particle with the occupation probability $(1-\rho)$. Therefore,
relabeling the single-site states does not lead to new models.}
\end{itemize}
Considering the above mentioned transformations, classification
can be done. Using the three parameters
$\displaystyle{\frac{a_1}{a_2}}\ ,\
\displaystyle{\frac{b_1}{b_2}}\ ,\
\displaystyle{\frac{(1-\rl)}{(1-\rr)}}$, and the two sets of
equations (\ref{constraint01}), two distinct models can be
obtained. The two cases are
\begin{eqnarray}
&&K_1)\ \ \frac{a_1}{a_2}=\frac{b_1}{b_2}<\frac{(1-\rl)}{(1-\rr)}\qquad\qquad \cr
&&K_2)\ \ \frac{a_1}{a_2}<\frac{b_1}{b_2}<\frac{(1-\rl)}{(1-\rr)}
\end{eqnarray}
So it is enough to study the cases $K_1$, and $K_2$.

\subsubsection {$K_1$}
Resulting from the condition
$\displaystyle{\frac{a_1}{a_2}=\frac{b_1}{b_2}}$, the non-zero
reaction rates are
\begin{eqnarray}
&&\{\omega_{23},\omega_{24},\omega_{27}\},\qquad\{\omega_{56},\omega_{58},\omega_{59}\},\cr
&&\{\omega_{32},\omega_{34},\omega_{37}\},\qquad\{\omega_{65},\omega_{68},\omega_{69}\},\cr
&&\{\omega_{42},\omega_{43},\omega_{47}\},\qquad\{\omega_{85},\omega_{86},\omega_{89}\},\cr
&&\{\omega_{72},\omega_{73},\omega_{74}\},\qquad\{\omega_{95},\omega_{96},\omega_{98}\},
\end{eqnarray}
in which particles A and B can convert to each other, but no
particle can annihilate to an empty state. Thus the total number
of empty sites and consequently the total number of particles are
conserved in the bulk. This model is symmetric under the the
transformation A$\leftrightarrow$B. As it will be seen the
relations we obtain reflect this symmetry. We have
\begin{eqnarray}
&(\omega_{56}+\omega_{58})a_1b_1-\omega_{+5}a_1^2+\omega_{59}b_1^2=0,&\cr
&(\omega_{96}+\omega_{98})a_1b_1+\omega_{95}a_1^2-\omega_{+9}b_1^2=0,&\cr\cr
&\displaystyle{\frac{\omega_{32}+\omega_{34}+\omega_{72}+\omega_{74}}{\omega_{23}+
\omega_{43}+\omega_{27}+\omega_{47}}=\frac{b_1}{a_1}},&
\end{eqnarray}
$d_1$ is given by
\begin{eqnarray}
d_1&=&\frac{a_1}{a_2}(\omega_{24}+\omega_{27}\frac{b_1}{a_1})=\frac{a_1}{a_2}(\omega_{37}+\omega_{34}\frac{a_1}{b_1})=\cr
&=&\frac{1-\rl}{1-\rr}(\omega_{42}+\omega_{43}\frac{b_1}{a_1})=\frac{1-\rl}{1-\rr}(\omega_{73}+\omega_{72}\frac{a_1}{b_1}),
\end{eqnarray}
$d_1$ and $d_2$ are related to each other through
\begin{eqnarray}
\frac{d_1}{(1-\rl)a_1}=\frac{d_2}{(1-\rr)a_2},
\end{eqnarray}
The boundary rates together with the reaction rates satisfy these
relations
\begin{eqnarray}
&&\rl((\omega_{32}+\omega_{72})a_1-(\omega_{23}+\omega_{43})b_1)=\cr\cr
&&=(-(\omega_{56}+\omega_{86}+\omega_{96})+\omega_{68})a_1b_1+\omega_{65}a_1^2+\omega_{69}b_1^2=\cr\cr
&&=(\alpha^\ell_{12}-\alpha^\ell_{13})a_1b_1-(-\alpha^\ell_{32}a_1+\alpha^\ell_{23}b_1)\rl,
\end{eqnarray}
$d_1$ and $d_2$ can be described by boundary rates through
\begin{eqnarray}
&&d_1=\frac{-(\alpha^\ell_{21}+\alpha^\ell_{31})(1-\rl)+\alpha^\ell_{12}a_1+\alpha^\ell_{13}b_1}{\rho_1-\rho_2},\cr\cr
&&d_2=\frac{(\alpha^r_{21}+\alpha^r_{31})(1-\rr)-\alpha^r_{12}a_2-\alpha^r_{13}b_2}{\rho_1-\rho_2},
\end{eqnarray}
and also
\begin{eqnarray}
\alpha^{\ell/r}_{21}b_1=\alpha^{\ell/r}_{31}a_1,
\end{eqnarray}
Defining $\alpha^{\ell+r}_{ij}:=\alpha^\ell_{ij}+\alpha^r_{ij}$, one arrives at
\begin{equation}
(\alpha^{\ell+r}_{12}-\alpha^{\ell+r}_{13})a_2=(-\alpha^{\ell+r}_{32}a_2+\alpha^{\ell+r}_{23}b_2)\frac{\rr}{b_2}
\end{equation}
And finally the rates $D'_1,D'_2$ are
\begin{eqnarray}
&&D'_1=d_2-d_1\frac{b_2}{b_1}+\frac{\alpha^\ell_{21}}{a_1}=
d_2\frac{a_2-a_1}{a_2}+\frac{\alpha^\ell_{12}a_1+\alpha^\ell_{13}b_1}{(1-\rl)\rl},\cr\cr
&&D'_2=d_1-d_2\frac{a_1}{a_2}+\frac{\alpha^r_{21}}{a_2}=d_1\frac{a_1-a_2}{a_1}+
\frac{\alpha^r_{13}b_2+\alpha^r_{12}a_2}{(1-\rr)\rr}.
\end{eqnarray}
\\

Setting all the transition rates  equal to zero except for the
diffusion rates $\omega_{73},\omega_{37},\omega_{42},\omega_{24}$ and the exchange
rates $\omega_{86},\omega_{68}$, the results of \cite{RS} can be obtained
 \begin{equation}
\omega_{68}=\omega_{86},\qquad\omega_{24}=\omega_{37}=d_1\frac{a_2}{a_1},\qquad\omega_{42}=\omega_{73}=d_1\frac{(1-\rr)}{(1-\rl)}
\end{equation}
 \begin{equation}
\dis\frac{d_1}{(1-\rl)a_1}=\frac{d_2}{(1-\rr)a_2}
\end{equation}

\begin{eqnarray}
&\dis\alpha^{\ell/r}_{21}b_1=\alpha^{\ell/r}_{31}a_1,&\cr\cr &
(\alpha^{\ell/r}_{12}-\alpha^{\ell/r}_{13})a_1b_1=(-\alpha^{\ell/r}_{32}a_1+\alpha^{\ell/r}_{23}b_1)\rho_1,&\cr\cr
&\dis
d_1=\frac{-\alpha^\ell_{31}(1-\rl)-\alpha^\ell_{32}a_1+(\alpha^\ell_{13}+\alpha^\ell_{23})b_1}{b_1-b_2}&\cr\cr
&\dis
d_2=\frac{\alpha^r_{31}(1-\rr)+\alpha^r_{32}a_2-(\alpha^r_{13}+\alpha^r_{23})b_2}{b_1-b_2}&
\end{eqnarray}
\begin{eqnarray}
&&D'_1=(\omega_{42}-\omega_{24})\rr+\frac{\alpha^\ell_{31}}{b_1}=
(\omega_{24}-\omega_{42})(1-\rr)+\frac{\alpha^\ell_{13}b_1+\alpha^\ell_{12}a_1}{(1-\rl)\rl},\cr\cr
&&D'_2=(\omega_{24}-\omega_{42})\rl+\frac{\alpha^r_{31}}{b_2}=(\omega_{42}-\omega_{24})(1-\rl)+
\frac{\alpha^r_{13}b_2+\alpha^r_{12}a_2}{(1-\rr)\rr}.
\end{eqnarray}
It is seen that the rates of diffusion for both particles A and B
are the same as obtained in  \cite{KJS} for a single-species
model. So ASEP is a special example of the case
$K_1$.\\

The shock will be immobile ($d_1=d_2=D'_1=D'_2=0$), if
\begin{eqnarray}\label{immobile1}
&&\omega_{24}\ ,\ \omega_{42}\ ,\ \omega_{27}\ ,\ \omega_{72}\ ,\ \omega_{34}\ ,\
\omega_{43}\ ,\ \omega_{37}\ ,\ \omega_{73}=0
\end{eqnarray}
These reactions are
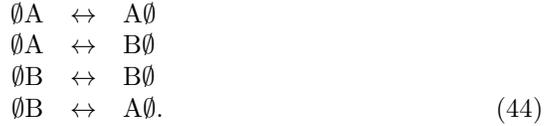
\begin{eqnarray}
\emptyset {\rm A}&\leftrightarrow& {\rm A}\emptyset\cr \emptyset
{\rm A}&\leftrightarrow&{\rm B} \emptyset\cr
 \emptyset{\rm B}&\leftrightarrow &{\rm B} \emptyset\cr
\emptyset{\rm B} &\leftrightarrow&{\rm A}\emptyset.
\end{eqnarray}
The above transition rates are responsible for particle
transportation from one site to another one. Whenever these rates
are nonzero, the shock position moves. Thus, to fix the shock
position, these rates should be zero. We also have

\begin{eqnarray}
&&\alpha^{\ell/r}_{12},\alpha^{\ell/r}_{13},\alpha^{\ell/r}_{21},\alpha^{\ell/r}_{31}=0
\end{eqnarray}
These boundary rates should be zero to maintain the total number
of particles in the bulk.
 So we have
\begin{eqnarray}
&\dis\frac{\omega_{32}+\omega_{74}}{\omega_{23}+\omega_{47}}=\frac{b_1}{a_1}&\cr
&\omega_{32}a_1-\omega_{23}b_1=\alpha^\ell_{32}a_1-\alpha^\ell_{23}b_1&\cr
&(\alpha^\ell_{32}+\alpha^r_{32})a_2=(\alpha^\ell_{23}+\alpha^r_{23})b_2&
\end{eqnarray}
The other relations remain unchanged.

\subsubsection{$K_2$}
In this model the occupation probabilities should satisfy the
following relation
\begin{eqnarray}\label{condition1}
&&\frac{(1-\rl)a_1}{b_1^2}=\frac{(1-\rr)a_2}{b_2^2}
\end{eqnarray}
Then the non-zero transition rates are
\begin{equation}
\omega_{24},\ \omega_{29},\ \omega_{42},\ \omega_{49},\ \omega_{92},\ \omega_{94},\
\omega_{37},\ \omega_{73},\ \omega_{68},\ \omega_{86}
\end{equation}
Therefore, besides the diffusion ($\omega_{24},\ \omega_{42}\ ,\ \omega_{37},\
\omega_{73}$) and exchange processes ($\omega_{68},\omega_{86}$), provided
that (\ref{condition1}) is satisfied, the processes $\emptyset
{\rm A}\leftrightarrow {\rm BB}$ and ${\rm
A}\emptyset\leftrightarrow {\rm BB}$ are also allowed. The
occupation probabilities should also satisfy the following
relation
\begin{eqnarray}
&&\dis\frac{a_1(1-\rr)-a_2(1-\rl)}{b_1(1-\rr)-b_2(1-\rl)}=\frac{a_1b_2-a_2b_1}{a_1(1-\rr)-a_2(1-\rl)}
\end{eqnarray}
This relation should be held in order to guarantee the possibility
of the simultaneous existence of the two uniform parts in the
bulk. Then we have
\begin{eqnarray}
&\omega_{37}=\omega_{86},&\cr
&\omega_{73}=\omega_{68},&\cr
&(\omega_{92}+\omega_{94})(1-\rl)a_1-(\omega_{29}+\omega_{49})b_1^2=0,&
\end{eqnarray}
$d_1$ and $d_2$ are connected through
\begin{eqnarray}
&&\frac{d_1}{(1-\rl)a_1}=\frac{d_2}{(1-\rr)a_2}
\end{eqnarray}
$d_1$ is given by
\begin{eqnarray}
d_1&=&\frac{a_1}{a_2}\omega_{37}=\frac{(1-\rl)}{(1-\rr)}\omega_{73}=\frac{a_1}{a_2}(\omega_{24}+\omega_{29}\frac{b_1}{a_1}\
\frac{a_1-a_2}{\rl-\rr})=\cr\cr
&=&\frac{(1-\rl)}{(1-\rr)}(\omega_{42}+\omega_{49}\frac{b_2}{a_2}\frac{a_1-a_2}{\rl-\rr})=
\frac{(1-\rl)a_1}{(a_1b_2-a_2b_1)}(\omega_{92}\frac{a_2}{b_2}-\omega_{94}\frac{a_1}{b_1})
\end{eqnarray}
Defining
$\alpha^\ell_{ij}+\alpha^r_{ij}=:\alpha^{\ell+r}_{ij}$, then $D'_1,D'_2$ will be
\begin{eqnarray}
&&D'_1=\frac{1}{b_2-b_1}(-\alpha^{\ell+r}_{31}(1-\rr)-\alpha^{\ell+r}_{32}a_2+(\alpha^{\ell+r}_{13}+\alpha^{\ell+r}_{23})b_2)\cr
&&D'_2=\frac{1}{b_2-b_1}(\alpha^{\ell+r}_{31}(1-\rl)+\alpha^{\ell+r}_{32}a_1-(\alpha^{\ell+r}_{13}+\alpha^{\ell+r}_{23})b_1)
\end{eqnarray}
The boundary relations have been included in Appendix.
A question which may arise is that: is there any solution for these set of conditions?
These conditions are some constraints on the reaction rates, and occupation probabilities of the particles, $a_i$ and  $b_i$, which should be checked. In addition to these equations, there are also some inequalities;  reaction rates should be nonnegative and   $0\leq a_i, b_i\leq 1$.
It is difficult to check all these analytically. We have solved numerically these set of equations and inequalities. It is seen that there are solutions for these set of equations. So all the conditions on the parameters can be fulfilled simultaneously.\\

This case does not have a similar one-species model. To have a
one-species analog, the processes in which all the three
single-site states are involved should vanish, i.e. these
processes $\rm A\emptyset\leftrightarrow BB$ should not happen.
This means that just the diffusion and exchange transition rates
can be held which leads to a contradiction to the inequality
$\frac{a_1}{a_2}<\frac{b_1}{b_2}<\frac{(1-\rl)}{(1-\rr)}$.

Moreover, immobile shock is not possible in this case. If we set
all the hopping rates of the shock position equal to zero, then
all the transition rates become accordingly zero. So, this model
does not admit immobile shocks.

\subsubsection{Generalization to more than two species }
Let us consider an $n$-species system. The local hamiltonian $h$,
acting on two adjacent sites, is an $(n+1)\times(n+1)$ matrix. We
assume a single-shock measure consisting of the single-site state
vectors $|u\rangle$ and $|v\rangle$. It can be seen from the
equations (\ref{hamiltinian},\ \ref{shock2}), if the equations
governing the evolution of $h|u\rangle\ot |u\rangle,\
h|u\rangle\ot |v\rangle$ and $h|v\rangle\ot |v\rangle$ are given,
the dynamics of the shock in the bulk of the lattice is fully
determined. To facilitate the calculation, we choose a new basis
like $\zeta=\{|u\rangle, |v\rangle,|x_1\rangle,\dots,
|x_{n-1}\rangle\}$, where $|x_{i}\rangle$'s are arbitrary vectors
ineffective in the final results. Adopting the same strategy as
Section 3, from expanding the three vectors $h|u\rangle\ot
|u\rangle, h|u\rangle\ot |v\rangle$ and $h|v\rangle\ot |v\rangle$
in the new basis and then substituting these expansions into
(\ref{shock2}), one obtains the following equations for the
uniform parts of the shock
\begin{eqnarray}\label{uniform}
&&h|u\rangle\ot|u\rangle=E_1(|u\rangle\ot|v\rangle-|v\rangle\ot|u\rangle)
+\sum_{\mu=1}^{n-1}E_{\mu+1}(|u\rangle\ot|x_{\mu}\rangle-|x_{\mu}\rangle\ot|u\rangle),\cr
&&h|v\rangle\ot|v\rangle=-F_1(|v\rangle\ot|u\rangle-|u\rangle\ot|v\rangle)
+\sum_{\mu=1}^{n-1}F_{\mu+1}(|v\rangle\ot|x_{\mu}\rangle-|x_{\mu}\rangle\ot|v\rangle),\qquad
\end{eqnarray}
and the following equation for the shock position
\begin{eqnarray}\label{shock}
&&h|u\rangle\ot|v\rangle=(d_2-F_1)(|u\rangle\ot|u\rangle-|u\rangle\ot
|v\rangle)+(d_1-E_1)(|v\rangle\ot|v\rangle-|u\rangle\ot|v\rangle)\cr
&&\quad\qquad\qquad+\sum_{\mu=1}^{n-1}F_{\mu+1}|u\rangle\ot|x_{\mu}\rangle-\sum_{\mu=1}^{n-1}E_{\mu+1}|x_{\mu}\rangle\ot
|v\rangle.\end{eqnarray} E's and F's are some constants depending
on the reaction rates and probabilities. When all the occupation
probabilities are different from zero and one, the equations
(\ref{uniform}) seem to be sufficient to determine the non-zero
rates. In this case, the equation (\ref{shock}) for the shock
position either manages to link the two uniform parts or
completely fails and rejects the validity of the whole model.

The equations (\ref{uniform}) are both in the following form
\begin{eqnarray}
&&h|w\rangle\ot|w\rangle=|w\rangle\ot|x\rangle-|x\rangle\ot|w\rangle
\end{eqnarray}
We define $|{\cal R}\rangle:= h|w\rangle\ot |w\rangle$ whose
elements are $R_k$'s. We have obtained some relations for these
elements by generalizing our previous knowledge of what these
relations are for one and two-species systems,
(\ref{constraint01}). We have
\begin{eqnarray}\label{R}
&&R_{(\xi-1)(n+1)+\xi}=0 \qquad,\xi=1,\dots,n+1
\end{eqnarray}
These relations belong to the two-site states in which either both
sites are empty or occupied by particles of the same type, for
example $\emptyset\emptyset$ or $\rm A \rm A$ . We also have

\begin{equation}\label{Rmirror}
R_{(\psi-1)(n+1)+\phi}+R_{(\phi-1)(n+1)+\psi}=0,\left\{
                                              \begin{array}{ll}
                                                \psi\neq \phi,  \\
                                                \psi=1,\cdots,n \\
                                                 \phi=2,\dots,n+1
                                              \end{array}
                                            \right.
\end{equation}

Each relation of this kind is related to those two-site states
which are the mirror of each other, for example $\emptyset\ A$ and
$A\ \emptyset$.

It is known that, in the case of single-species models there is
only one model with $0<\rho<1$, admitting shock as a random walker
\cite{KJS}. This model is the asymmetric simple exclusion process
(ASEP). For the two-species case, we have found two models here.
One of these models is a generalization of ASEP, and the other one
has no single-species analog.

Let us consider a three-species model. The single-site states
respectively are empty site denoted by $\emptyset$ and occupied
sites by {\rm A}, {\rm B} and {\rm C}. Then the two-site states
become:
\begin{eqnarray}
1)\ \emptyset\emptyset\quad\  & 5)\  {\rm A}\emptyset\ \quad & 9)\
\ {\rm B}\emptyset\qquad\  13)\ {\rm C} \emptyset \cr 2)\
\emptyset {\rm A} \quad & 6)\  {\rm A}{\rm A}\quad & 10)\ {\rm
B}{\rm A}\qquad 14)\ {\rm C}{\rm A}\cr 3)\ \emptyset{\rm B} \quad
& 7)\ {\rm A}{\rm B}\quad & 11)\  {\rm B}{\rm B}\qquad 15)\ {\rm
C}{\rm B}\cr 4)\ \emptyset{\rm C}\quad & 8)\ {\rm A}{\rm C}\quad &
12)\ {\rm B}{\rm C}\qquad 16)\ {\rm C}{\rm C}
\end{eqnarray}
and the equations (\ref{R}) and (\ref{Rmirror}) give
\begin{eqnarray}
&&R_1=R_6=R_{11}=R_{16}=0\cr\cr &&R_2+R_5=0,\quad\
R_3+R_9=0,\quad\ R_4+R_{13}=0\cr\cr &&R_7+R_{10}=0,\quad
R_8+R_{14}=0,\quad R_{12}+R_{15}=0
\end{eqnarray}
To do classification, one should compare the following quantities
with each other
\begin{eqnarray}\label{ratio}
&&\frac{(1-\rho_1)^2}{(1-\rho_2)^2}\ ,\quad\frac{a_1^2}{a_2^2}\
,\quad\frac{b_1^2}{b_2^2}\ ,\quad\frac{c_1^2}{c_2^2}\ ,
\quad\frac{(1-\rho_1)a_1}{(1-\rho_2)a_2}\ ,\cr\cr\cr
&&\frac{(1-\rho_1)b_1}{(1-\rho_2)b_2}\
,\quad\frac{(1-\rho_1)c_1}{(1-\rho_2)c_2}\
,\quad\frac{a_1b_1}{a_2b_2}\ ,\quad\frac{a_1c_1}{a_2c_2}\
,\quad\frac{b_1c_1}{b_2c_2}\ .
\end{eqnarray}
where $a,\ b$ and $c$ are the occupation probabilities of
particles and $\rho$ is defined as ($\rho:=a+b+c$). These
parameters are suitable criteria for the ratio of the number of
the two-site states of the two uniform parts. As we know, how much
greater the number of a definite two-site state is, the associated
process is more likely to happen. Therefore, the reason for which
the parameters (\ref{ratio}) are helpful to introduce tractable
models is that they can be used to control the processes in the
lattice. Since the parameters (\ref{ratio}) are the square or
product of the parameters $\frac{a_1}{a_2}\ ,\ \frac{b_1}{b_2}\ ,\
\frac{c_1}{c_2}$ and $\frac{(1-\rl)}{(1-\rr)}$, it will be easier
to first compare
these parameters.\\

Three distinct models will be obtained for a three species
lattice.
\begin{itemize} \item{The cases with one inequality like
\begin{equation}\label{case1}
\frac{a_1}{a_2}=\frac{b_1}{b_2}=\frac{c_1}{c_2}<\frac{(1-\rl)}{(1-\rr)}
\end{equation}
For this case, the non-zero transition rates are
\begin{eqnarray}
& \omega_{ij}, \quad &i\neq j,\quad i,j=6, 7, 8, 10, 11, 12, 14, 15,
16, \cr & \omega_{lk}, \quad &l\neq k,\quad  l,k=2 ,3 ,4 ,5 , 9 ,13.
\end{eqnarray}
in which particles A, B and C can convert to each other, but the
number of empty sites is conserved. Resulting from (\ref{case1}),
this model is invariant under the transformations $\rm
A\leftrightarrow \rm B$, $\rm A\leftrightarrow \rm C$ and $\rm
C\leftrightarrow \rm B$. It shows that, from the four single-site
states of this case, the states A, B and C are similar to each
other. Another model of this type is
\begin{equation}
\frac{a_1}{a_2}=\frac{b_1}{b_2}<\frac{c_1}{c_2}=\frac{(1-\rl)}{(1-\rr)}
\end{equation}
The non-zero transition rates are
\begin{eqnarray}
&&\omega_{ef}, \quad e\neq f,\quad e,f=2, 3, 5, 8, 9 , 12, 14, 15,\cr
&&\omega_{gh},\quad g\neq h,\quad g,h=6, 7, 10, 11 ,\cr &&\omega_{op},
\quad o\neq p,\quad o,p=1 , 4, 13, 16 .
\end{eqnarray}
This model has the symmetries $\rm A\leftrightarrow \rm B$ and
$\rm C\leftrightarrow \emptyset$, so the states A with B and empty
site with C are similar to each other. Therefore, for the cases
with one inequality, one expects an analogous single-species
model, resembling two reduction in the number of particles.}

\item{The cases with two inequalities like
\begin{equation}
\frac{a_1}{a_2}=\frac{b_1}{b_2}<\frac{c_1}{c_2}<\frac{(1-\rl)}{(1-\rr)}
\end{equation}
This model has the symmetry $\rm A\leftrightarrow \rm B$, thus it
is expected to have a similar two-species model, resembling one
reduction in the number of particles. The nonzero transition rates
are
\begin{eqnarray}
&\qquad\qquad &\omega_{4,13}\ ,\ \omega_{13,4}\cr\cr &\omega_{i'j'}, \quad
&i'\neq j',\quad\ i',j'= 6, 7, 10, 11,\cr &\omega_{l',k'},\quad
&l'\neq k',\quad\ l',k'= 8, 12, 14, 15,\cr &\omega_{e'f'}, \quad
&e'\neq f',\quad\ e',f'=2 , 3, 5, 9, 16 .
\end{eqnarray}
in which, besides the diffusion and exchange processes, particles
A and B can convert to each other and the processes ($\emptyset\rm
A,\ \rm A\emptyset,\ \emptyset\rm B,\ \rm
B\emptyset\leftrightarrow \rm C\rm C$) are also possible provided
that the following relation is satisfied
\begin{eqnarray}
\frac{(1-\rl)a_1}{(1-\rr)a_2}=\frac{c_1^2}{c_2^2}
\end{eqnarray}
 There is another essential
 condition for the occupation probabilities which is
\begin{eqnarray}
&&((a_1+b_1)(1-\rr)-(a_2+b_2)(1-\rl))^2=\cr\cr&&
=(c_1(1-\rr)-c_2(1-\rl))((a_1+b_1)c_2-(a_2+b_2)c_1)
\end{eqnarray}
This relation comes from the equation (\ref{shock}) for the shock
position.
 }

\item{The cases with three inequalities like
\begin{equation}
\frac{a_1}{a_2}<\frac{b_1}{b_2}<\frac{c_1}{c_2}<\frac{(1-\rl)}{(1-\rr)}
\end{equation}
are expected to be originally three-species models. The nonzero
transition rates are
\begin{eqnarray}
&& \omega_{7,10}\ ,\ \omega_{10,7}\ ,\ \omega_{4,13}\ ,\ \omega_{13,4}\ ,\cr\cr
&&\omega_{g'h'}, \qquad g'\neq h',\quad g',h'= 2, 5, 12, 15,\cr
&&\omega_{o'p'}, \qquad o'\neq p',\quad o',p'= 8, 11, 14,\cr
&&\omega_{q's'}, \qquad q'\neq s',\quad q',s'= 3, 9, 16 .
\end{eqnarray}
with the conditions
\begin{equation}
\frac{(1-\rl)a_1}{(1-\rr)a_2}=\frac{b_1c_1}{b_2c_2},\qquad\frac{(1-\rl)b_1}{(1-\rr)b_2}=\frac{c_1^2}{c_2^2},\qquad\frac{a_1c_1}{a_2c_2}=\frac{b_1^2}{b_2^2}
\end{equation}
In this case, besides the diffusion and exchange processes,
resulting from the above relations, the processes ($\emptyset\rm
A,\ \rm A\emptyset\leftrightarrow \rm B\rm C,\ \rm C\rm B$),
($\emptyset\rm B,\ \rm B\emptyset\leftrightarrow \rm C\rm C$) and
($\rm A\rm C,\ \rm A\rm C\leftrightarrow \rm B\rm B$)
 are also allowed. The
occupation probabilities should also satisfy another relation
coming from the equation (\ref{shock}).}

\end{itemize}
Accordingly for $n$-species systems, it is expected to find $n$
distinct models. One for which there is a similar single-species
model, one with a two-species analog, $\cdots$, and finally there
is an $n$-species model which has no $k$-species analog with
$k<n$.

\subsection {shocks containing the occupation probabilities 0 and 1}
Now, we study the shocks in which some parts of the lattice is empty (or equivalently completely occupied).
\subsubsection { $\rl\neq0,1,\rr=0$}
Assume $a_2=b_2=0$ and $a_1,b_1\neq0,1$. In this case, $\omega_{i1}=0$
to prevent particle production in the empty part of the shock, and
subsequently $\omega_{1i}=0$, otherwise all the two-site states can
convert to $\emptyset\emptyset$ but there is no way to leave this
state. All the other remaining rates can be non-zero. Besides the
relations (\ref{constraint01}) for $\rl,b_1,a_1$, the other
relations are as follows. The parameters $d_1,d_2$ are given by
\begin{eqnarray}
d_1&=&\frac{1}{a_1}[(\omega_{42}a_1+\omega_{43}b_1)(1-\rl)+(\omega_{46}+\omega_{48})a_1b_1+\omega_{45}a_1^2+\omega_{49}b_1^2]\cr
&=&\frac{1}{b_1}[(\omega_{72}a_1+\omega_{73}b_1)(1-\rl)+(\omega_{76}+\omega_{78})a_1b_1+\omega_{75}a_1^2+\omega_{79}b_1^2],\cr&&\cr
d_2&=&\frac{1}{(1-\rl)
a_1}(\omega_{24}a_1+\omega_{27}b_1)=\frac{1}{(1-\rl)
b_1}(\omega_{34}a_1+\omega_{37}b_1)\cr&=&\frac{1}{a_1^2}(\omega_{54}a_1+\omega_{57}b_1)=
\frac{1}{a_1b_1}(\omega_{64}a_1+\omega_{67}b_1)\cr&=&\frac{1}{b_1a_1}(\omega_{84}a_1+
\omega_{87}b_1)=\frac{1}{b_1^2}(\omega_{94}a_1+\omega_{97}b_1).
\end{eqnarray}
The boundary rates together with the reaction rates satisfy these
relations
\begin{eqnarray}
&&(\alpha^\ell_{12}-\alpha^\ell_{13})a_1b_1+(\alpha^\ell_{32}a_1-\alpha^\ell_{23}b_1)\rl=
(-\omega_{+4}a_1+\omega_{47}b_1+d_2a_1)\rl=\cr\cr &&
=[(\omega_{62}+\omega_{64})a_1+(\omega_{63}+\omega_{67})b_1](1-\rl)+\omega_{65}a_1^2+\omega_{69}b_1^2\cr
&&\quad+(-\omega_{+6}+\omega_{68})a_1b_1,
\end{eqnarray}
and we have
\begin{eqnarray}
-d_2\rl(1-\rl)+d_1\rl
=-(\alpha^\ell_{21}+\alpha^\ell_{31})(1-\rl)+\alpha^\ell_{12}a_1+\alpha^\ell_{13}b_1,
\end{eqnarray}
Particles should not enter
 the right boundary, so $\alpha^r_{31}=\alpha^r_{21}=0$. The relations between the boundary
rates are
\begin{eqnarray}
&&\alpha^\ell_{21}b_1=\alpha^\ell_{31}a_1.
\end{eqnarray}
Using the definition $\alpha^{\ell+r}_{ij}:=\alpha^\ell_{ij}+\alpha^r_{ij}$, one arrives at
\begin{eqnarray}
(\alpha^{\ell+r}_{12}-\alpha^{\ell+r}_{13})a_1b_1=(-\alpha^{\ell+r}_{32}a_1+\alpha^{\ell+r}_{23}b_1)\rl.
\end{eqnarray}
$D'_1$, $D'_2$ are
\begin{eqnarray}
&&D'_1=\frac{\alpha^\ell_{31}}{b_1}\cr
&&D'_2=d_1-d_2(1-\rl)+\frac{\alpha^r_{12}a_1+\alpha^r_{13}b_1}{\rl}.
\end{eqnarray}
The shock will be immobile if all the transition rates from the
fourth and seventh rows and columns of $h$ except for
$\omega_{47},\omega_{74}$ become zero, i.e. the processes starting from
(ending to) the states $\rm A\emptyset,\ \rm B\emptyset$ which
make the shock position move to the right (left). Also the
following rates at the boundaries should be zero
\begin{equation}
\alpha^\ell_{21}=\alpha^\ell_{31}=\alpha^\ell_{12}=\alpha^\ell_{13}=\alpha^r_{12}=\alpha^r_{13}=0.
\end{equation}
which means no particle enter or leave the boundaries.
The other relations will remain unchanged.

\subsubsection{ $\rl=1,\rr=0$}
Consider $a_2=b_2=0$ and $a_1+b_1=1$. We assume  $a_1,b_1\neq0,1$
to have a two species model. The non-zero transition rates are
\vskip0.5cm\hskip1.5cm
\begin{tabular}{ccccccccc}&$\omega_{12}$&$\omega_{13}$&$\omega_{14}$&&&$\omega_{17}$&&\\   &&$\omega_{23}$&&&&&&\\
&$\omega_{32}$&&&&&&&  \\&$\omega_{42}$&$\omega_{43}$&&&&$\omega_{47}$&&  \\&$\omega_{52}$&$\omega_{53}$&$\omega_{54}$&&$\omega_{56}$&$\omega_{57}$&$\omega_{58}$&$\omega_{59}$
\\&$\omega_{62}$&$\omega_{63}$&$\omega_{64}$&$\omega_{65}$&&$\omega_{67}$&$\omega_{68}$&$\omega_{69}$  \\&$\omega_{72}$&$\omega_{73}$&$\omega_{74}$&&&&&
\\&$\omega_{82}$&$\omega_{83}$&$\omega_{84}$&$\omega_{85}$&$\omega_{86}$&$\omega_{87}$&&$\omega_{89}$
\\&$\omega_{92}$&$\omega_{93}$&$\omega_{94}$&$\omega_{95}$&$\omega_{96}$&$\omega_{97}$&$\omega_{98}$&
\end{tabular}
\vskip0.5cm \noindent There are no constraints on $\omega_{i2}$ and
$\omega_{i3}$. One should expect it,  because the states
($\emptyset$A) and ($\emptyset$B) do not exist in the final state.
We also have
\begin{eqnarray}
&&(\omega_{56}+\omega_{58})a_1b_1-\omega_{+5}a_1^2+\omega_{59}b_1^2=0\cr
&&(\omega_{96}+\omega_{98})a_1b_1+\omega_{95}a_1^2-\omega_{+9}b_1^2=0.
\end{eqnarray}
$d_1,d_2$ are given by
\begin{eqnarray}
d_1&=&\omega_{14}a_1+\omega_{17}b_1\cr
d_2&=&\frac{1}{a_1^2}(\omega_{54}a_1+\omega_{57}b_1)=\frac{1}{a_1b_1}(\omega_{64}a_1+\omega_{67}b_1)\cr
&=&\frac{1}{b_1a_1}(\omega_{84}a_1+\omega_{87}b_1)=\frac{1}{b_1^2}(\omega_{94}a_1+\omega_{97}b_1).
\end{eqnarray}
The following relations should also be satisfied by reaction rates
\begin{eqnarray}
\alpha^\ell_{32}a_1-\alpha^\ell_{23}b_1&=&(-\omega_{+6}+\omega_{68})a_1b_1+\omega_{65}a_1^2+
\omega_{69}b_1^2\cr
&=&(d_2+d_1)a_1-\omega_{+4}a_1+\omega_{47}b_1.
\end{eqnarray}
The left part of the shock should be fully occupied, so
$\alpha^\ell_{12}=\alpha^\ell_{13}=0$ and the right part of the
shock should be empty, thus $\alpha^r_{21}=\alpha^r_{31}=0$. The
relations between the boundary rates are
\begin{eqnarray}
&\alpha^\ell_{21}b_1=\alpha^\ell_{31}a_1&\cr\cr
&(\alpha^r_{12}-\alpha^r_{13})a_1b_1=-(\alpha^r_{32}+\alpha^\ell_{32})a_1+(\alpha^r_{23}+\alpha^\ell_{23})b_1&
\end{eqnarray}
$D'_1, D'_2$ are given by
\begin{eqnarray}
&&D^{'}_1=\frac{\alpha^\ell_{31}}{b_1}\cr
&&D^{'}_2=\alpha^r_{12}a_1+\alpha^r_{13}b_1.
\end{eqnarray}
 The shock will be immobile if we eliminate the processes
 ($\rm A\emptyset,\ \rm B\emptyset\rightarrow\emptyset\emptyset$), i.e. $\omega_{14}=\omega_{17}=0$, to prevent the shock position from moving to the left,
 and also the processes ($\rm A\emptyset,\ \rm B\emptyset\rightarrow \rm A\rm B,\ \rm B\rm A,\ A\rm A,\ \rm B\rm
 B$) should vanish to prevent the shock position from moving to the
right, i.e the transition rates $\omega_{47},\omega_{74}$ are the only
non-zero rates of the forth and seventh columns of
 $h$.
 The following boundary rates should be also zero
\begin{eqnarray}
&&\alpha^\ell_{21}=\alpha^\ell_{31}=\alpha^r_{12}=\alpha^r_{13}=0.
\end{eqnarray}
which means no particle enter or leave the boundaries.

Some other shapes of the shocks for two species systems can be
also predicted
\begin{eqnarray}
&&(a_1=0,\qquad a_2\neq0,1,\quad b_1\neq0,1,\quad b_2=0)\cr
&&(a_1=0,\qquad a_2\neq0,1,\quad b_1=1,\qquad b_2=0)\cr
&&(a_1\neq0,1,\quad a_2\neq0,1,\quad b_1\neq0,1, \quad b_2=0,\
a_1+b_1\neq0,1)\cr &&(a_1\neq0,1,\quad a_2\neq0,1,\quad
b_1\neq0,1,\quad b_2=0,\ a_1+b_1=1)
\end{eqnarray}
It can be shown that the models with $a_1=a_2,\ (a_1, a_2\ne
0,1)$, $b_1\ne b_2,\ (b_1, b_2\ne 0,1)$, and $\rho_1, \rho_2\ne
0,1$, do not exist.

\section{Dynamical phase transition for two-species models }
In this section dynamical phase transition for the three cases
$K_1$, $(\rho_1\ne 0,1, \rho_2=0)$, and $(\rho_1=1, \rho_2=0)$ is
studied. In \cite{AA}, phase transition in single-species models
possessing shock solutions is studied. It will be shown that there
are three phases. For some region of the parameter space, the
relaxation time is independent of the reaction rates at the
boundaries. Changing continuously the reaction rates at the
boundaries, there is a point where the relaxation time begins
changing, so that at this point there is a jump in the derivative
of the relaxation time with respect to the reaction rates at
boundaries. This is the dynamical phase transition. So the
dynamical phase transition studied here is a discontinuity in the
derivative of the relaxation time (from zero to nonzero) with
respect to reaction rates in the bulk and at the boundaries. We
use the method presented there, for our two-species models.
Defining the two parameters ${\cal A}$, and ${\cal B}$ as
\begin{eqnarray}
&{\cal A}:=\displaystyle{\frac{d_1-D'_2}{\sqrt{d_1d_2}}},& {\cal
B}:=\frac{d_2-D'_1}{\sqrt{d_1d_2}},
\end{eqnarray}
it is shown in \cite{AA} depending on the phase of the system the
relaxation times of the models may  be $T_0$, $T_{\cal A}$ or
$T_{\cal B}$
\begin{equation}
T_0=\frac{1}{(\sqrt d_1-\sqrt d_2)^2},\, T_{\cal
A}=\displaystyle{\frac{d_1-D'_2}{D'_2(d_1-D'_2-d_2)}},\, T_{\cal
B}=\frac{d_2-D'_1}{D'_1(d_2-D'_1-d_1)}.
\end{equation}
Three phases may occur,
\begin{eqnarray}
1)& {\cal A}<1,{\cal B}<1:& T_0,\cr 2)& {\cal A}<1,{\cal B}>1:&
T_{\cal B},\cr 3)& {\cal A}>1,{\cal B}<1:& T_{\cal A}.
\end{eqnarray}
 The case ${\cal A}>1,{\cal B}>1$ do not occur in none of the above mentioned models.

 Let us first consider the case $K_1$. we have
\begin{eqnarray}
&&{\cal
A}=\frac{(\omega_{42}+\omega_{43}\frac{b_1}{a_1})-\frac{\alpha^r_{21}}{a_2}}{\sqrt{(\omega_{24}+\omega_{27}
\frac{b_1}{a_1})(\omega_{42}+\omega_{43}\frac{b_1}{a_1})}}
=\frac{(\omega_{24}+\omega_{27}\frac{b_1}{a_1})-\frac{\alpha^r_{13}b_2+\alpha^r_{12}a_2}{(1-\rr)\rr}}
{\sqrt{(\omega_{24}+\omega_{27}\frac{b_1}{a_1})(\omega_{42}+\omega_{43}\frac{b_1}{a_1})}}\cr\cr
&&{\cal
B}=\frac{(\omega_{24}+\omega_{27}\frac{b_1}{a_1})-\frac{\alpha^\ell_{21}}{a_1}}{\sqrt{(\omega_{24}+\omega_{27}
\frac{b_1}{a_1})(\omega_{42}+\omega_{43}\frac{b_1}{a_1})}}
=\frac{(\omega_{42}+\omega_{43}\frac{b_1}{a_1})-\frac{\alpha^\ell_{13}b_1+\alpha^\ell_{12}a_1}{(1-\rl)\rl}}
{\sqrt{(\omega_{24}+\omega_{27}\frac{b_1}{a_1})(\omega_{42}+\omega_{43}\frac{b_1}{a_1})}}
\nonumber\end{eqnarray} If
$(\omega_{24}+\omega_{27}\frac{b_1}{a_1})>(\omega_{42}+\omega_{43}\frac{b_1}{a_1})$
one arrives at
\begin{equation}
{\sqrt{(\omega_{24}+\omega_{27}\frac{b_1}{a_1})(\omega_{42}+\omega_{43}\frac{b_1}{a_1})}}>(\omega_{42}
+\omega_{43}\frac{b_1}{a_1}).
\end{equation}
So
${\sqrt{(\omega_{24}+\omega_{27}\frac{b_1}{a_1})(\omega_{42}+\omega_{43}\frac{b_1}{a_1})}}>(\omega_{42}
+\omega_{43}\frac{b_1}{a_1})- \frac{\alpha^r_{21}}{a_2}$, which gives
${\cal A}<1$. For
\begin{eqnarray}
(\omega_{24}+\omega_{27}\frac{b_1}{a_1})>(\omega_{42}+\omega_{43}\frac{b_1}{a_1}),
\end{eqnarray}
it can be shown that ${\cal B}$ is also less than one. If
\begin{eqnarray}
(\omega_{24}+\omega_{27}\frac{b_1}{a_1})<(\omega_{42}+\omega_{43}\frac{b_1}{a_1}),
\end{eqnarray}
one again can  show that both ${\cal A}$, and ${\cal B}$ are less
than one. Thus this model belongs to the region 1. So the model
$K_1$ possesses  no dynamical phase transition, and the relaxation
time is $T_0$. $T_0$  has no dependence on boundary rates and only
depends on the reaction rates in the bulk. This result is the same
as the one found for ASEP, which is a single-species model. It
should be noted that this model is a generalization of ASEP. If
one forgets about the difference of two particles, $K_1$ changes
to ASEP.
\\

In the case of  $\rl\neq0,1,\rr=0$, defining
\begin{eqnarray}
&&\Omega_1:=\frac{1}{(1-\rl)}\big[(\omega_{42}+\omega_{43}\frac{b_1}{a_1})(1-\rl)+
(\omega_{46}+\omega_{48})b_1+\omega_{45}a_1+\omega_{49}\frac{b_1^2}{a_1}\big],\cr
&&\Omega_2:=(\omega_{24}+\omega_{27}\frac{b_1}{a_1}),
\nonumber\end{eqnarray}
one arrives at
\begin{eqnarray}
&&{\cal
A}=\frac{\Omega_2-\displaystyle{\frac{\alpha^r_{12}a_1+\alpha^r_{13}b_1}{\rl}}}{\sqrt{\Omega_1\Omega_2}}\cr
&&\cr &&{\cal
B}=\frac{\Omega_1-\displaystyle{\frac{\alpha^\ell_{12}a_1+\alpha^\ell_{13}b_1}{\rl(1-\rl)}}}{\sqrt{\Omega_1\Omega_2}}
=\frac{\displaystyle{\frac{\Omega_2}{(1-\rl)}-\frac{\alpha^\ell_{31}}{b_1}}}{\sqrt{\Omega_1\Omega_2}}
\end{eqnarray}
if $\Omega_1<\Omega_2$, then  ${\cal B}<1$ and ${\cal A}$ can be
smaller or greater than one and if $\Omega_1>\Omega_2$ ${\cal
A}<1$ and ${\cal B}$ can be smaller or greater than one. So, in
this case the regions 1,2 or 3 are possible. There are three
distinct relaxation times. $T_0$ is bulk dominated and depends
only on reaction rates in the bulk, while $T_{\cal A}$ and
$T_{\cal B}$ depend both on the reaction rates in the bulk, and at
the boundaries. So in this model there are three distinct phases
available for the system, and this model may experience dynamical
phase transitions. Changing the reaction rates at the bulk or at
the boundaries may lead to phase transitions.

And finally, in the case $\rl=1,\rr=0$, defining
\begin{eqnarray}
&&\Omega'_1:=(\omega_{14}a_1+\omega_{17}b_1),\cr
&&\Omega'_2:=(\omega_{54}+\omega_{64}+\omega_{84}+\omega_{94})a_1+(\omega_{57}+\omega_{67}+\omega_{87}+\omega_{97})b_1,
\end{eqnarray}
it can be shown that this model's parameters may be in any  of the
regions 1, 2 or 3. So this model may also experience dynamical
phase transitions.
It is known that single-species models have three distinct phases\cite{AA}. Here, It is seen that the two-species models considered here  are also restricted to the regions 1,2, and 3.

\section{Discussion}
In this work, we presented a method for classifying $n$-species
particle systems, possessing single shock solutions. This
classification provides a simple strategy, adopting only the
occupation probabilities of particles. We applied this method to
two- and then three-species systems.

We have made a detailed study on the four two-species models
$K_1$, $K_2$, ($\rl\neq0,1,\rr=0$) and ($\rl=1,\rr=0$). For all
these models some new results have been found here. In \cite{JM}
the three models $K_1$, ($\rl\neq0,1,\rr=0$) and ($\rl=1,\rr=0$)
have been studied. In this article, with the same assumptions,
more general solutions have been found. This shows there are some
unnecessary constraints in \cite{JM}, which  could be eliminated.
For example, for ($\rl=1,\rr=0$), there should be no constraints
on $\omega_{i2}$, $\omega_{i3}$, because the states $\emptyset A$,
$\emptyset B$ do not exist in the final state, and also the
transition rates $\omega_{56},\ \omega_{58},\ \omega_{59},\ \omega_{65},\
\omega_{68},\ \omega_{69},\ \omega_{85},\ \omega_{86},\ \omega_{89},\ \omega_{95},\
\omega_{96},\ \omega_{98}$ from the 5th, 6th, 8th, 9th columns of the
local hamiltonian $h$ were set equal to zero, while generally they
could be nonzero (see subsection 4.2.2).

The model investigated in \cite{TS2} resembles the model $K_2$ of
this paper in some of the initial assumptions; the nonzero bulk
transition rates and the condition (\ref{condition1}) of this
paper are the same. However, each paper has been presented a
completely distinct solution. In \cite{TS2}, the occupation
probabilities have been assumed to be equal to 0 and 1 in one part
of the lattice, which required three of the ten transition rates
to be zero. This model is a particular example of the more general
model $\rl\neq0,1,\rr=0$ (Subsection 4.2.1). We have found another
solution with the assumption that the occupation probabilities
should be necessarily different from 0 and 1. Consequently, one
should note that the two papers presented two different models.

Moreover, the model studied in Subsection 4.1.1 with a special
choice of symmetry has been previously studied in \cite{TS1}. A
special example of this model when the reaction rates are
eliminated has also been
presented in \cite{RS}. We derived the same relations for this case in Subsection 4.1.1.\\

\textbf{Acknowledgement}:  A.A. was partially supported by the
research council of the Alzahra University.

\section{Appendix}
The simplest forms of the boundary relations of the case $K_2$ are
\begin{eqnarray}
1)&&-\alpha^{\ell}_{21}(1-\rr)+(\alpha^\ell_{12}+\alpha^\ell_{32})a_2-\alpha^\ell_{23}b_2+\cr
&&(\alpha^\ell_{31}(1-\rr)+\alpha^\ell_{32}a_2-(\alpha^\ell_{13}+\alpha^\ell_{23})b_2)\frac{a_2-a_1}{b_2-b_1}=I_1\cr
2)&&-\alpha^r_{21}(1-\rr)+(\alpha^r_{12}+\alpha^r_{32})a_2-\alpha^r_{23}b_2+\cr
&&(\alpha^r_{31}(1-\rr)+\alpha^r_{32}a_2-(\alpha^r_{13}+\alpha^r_{23})b_2)\frac{a_2-a_1}{b_2-b_1}=-I_1\cr
3)&&-\alpha^{\ell}_{21}(1-\rl)+(\alpha^\ell_{12}+\alpha^\ell_{32})a_1-\alpha^\ell_{23}b_1+\cr
&&(\alpha^\ell_{31}(1-\rl)+\alpha^\ell_{32}a_1-(\alpha^\ell_{13}+\alpha^\ell_{23})b_1)\frac{a_2-a_1}{b_2-b_1}=I_2\cr
4)&&-\alpha^r_{21}(1-\rl)+(\alpha^r_{12}+\alpha^r_{32})a_1-\alpha^r_{23}b_1+\cr
&&(\alpha^r_{31}(1-\rl)+\alpha^r_{32}a_1-(\alpha^r_{13}+\alpha^r_{23})b_1)\frac{a_2-a_1}{b_2-b_1}=-I_2\cr
5)&&\frac{1}{b_2-b_1}(-\alpha^r_{31}(1-\rr)-\alpha^r_{32}a_2+(\alpha^r_{13}+\alpha^r_{23})b_2)=I_3\cr
6)&&\frac{1}{b_2-b_1}(\alpha^\ell_{31}(1-\rl)+\alpha^\ell_{32}a_1-(\alpha^\ell_{13}+\alpha^\ell_{23})b_1)=I_4
\end{eqnarray}
Defining
\begin{equation}
M:=\left(\frac{(a_1b_2-a_2b_1)(\rl-\rr)}{(a_1-a_2)(b_1-b_2)}\right)
\end{equation}
$I_1,\ I_3,\ I_2,\ I_4$ become
\begin{eqnarray}
&& I_1b_1=-I_2b_2\cr\cr
&&I_2=d_1\frac{(a_1b_2-a_2b_1)^3}{a_1a_2b_1}\
\frac{(\rl-\rr)}{(a_1-a_2)(b_1-b_2)}\cr\cr
&&I_4=d_1(1+M^2\frac{(1-2b_1)}{a_1(1-\rl)})\cr\cr &&I_3=d_1(\ \
\frac{b_2^2}{b_1^2}\ +M^2\frac{(1-2b_2)}{a_1(1-\rl)}).
\end{eqnarray}
As it can be seen, if one of the constants $d_1,\ I_1,\ I_3,\
I_2,\ I_4$ is given, the other constants will be obtained from the
above mentioned relations.

\end{document}